\documentclass[12pt]{article}
\usepackage{amsmath, amssymb, epsfig, natbib}
\usepackage{amsmath, bbm, amssymb}
\usepackage{amsthm}
\usepackage{comment}
\usepackage{tabularx}
\usepackage{graphicx, graphics,color}
\usepackage{array}
\usepackage{caption}
\usepackage{booktabs}
\usepackage{float} 
\usepackage{soul}  
\usepackage{url}

\usepackage{pdflscape}
\usepackage{enumitem, xcolor}
\usepackage{multirow}
\usepackage{hyperref}
\hypersetup{
	colorlinks=true,        
	citecolor=blue,         
	linkcolor=blue,         
	urlcolor=black,          
	bookmarksnumbered=true,  
	bookmarksopen=true,     
	breaklinks=true,        
	pdfstartview=FitH
}
\usepackage{epstopdf, mathrsfs, framed}
\usepackage{mathtools}
\usepackage{diagbox}
\usepackage[section]{placeins}
\colorlet{linkequation}{blue}
\usepackage{algorithmic}
\usepackage[ruled,linesnumbered,vlined]{algorithm2e}
\usepackage{cancel}
\usepackage{xr}
\newtheorem{corollary}{Corollary}

\newtheorem{proposition}{Proposition}
\newtheorem{example}{Example}

\usepackage{algorithmic}
\usepackage{thmtools}
\usepackage{xpatch}

\def\prob {{\rm Pr}}
\def\E{\mbox{\rm E}}
\def\var{\mbox{\rm var}}
\def\cov{\mbox{\rm cov}}
\def\sd{\mbox{\rm sd}}

\renewcommand{\vec}[1]{\mbox{\boldmath ${#1}$}}

\def\vmu{\vec{\mu}}

\date{}

\title{Consistency assessment and regional sample size calculation for MRCTs under random effects model}

\author{Xinru Ren$^{1}$ and Jin Xu$^{1,2,\ast}$}

\begin{document}
\maketitle

\noindent 
$^1$ School of Statistics, East China Normal University, Shanghai, China\\
$^2$ Key Laboratory of Advanced Theory and Application in Statistics and Data Science - MOE, East China Normal University, Shanghai, China\\
$^\ast$ Correspondence to: Jin Xu, Email: {jxu@stat.ecnu.edu.cn}

\begin{abstract}
Multi-regional clinical trials (MRCTs) have become common practice for drug development and global registration.  Once overall significance is established, demonstrating regional consistency is critical for local health authorities. Methods for evaluating such consistency and calculating regional sample sizes have been proposed based on the fixed effects model using various criteria. To better account for the heterogeneity of treatment effects across regions, the random effects model naturally arises as a more effective alternative for both design and inference. In this paper, we present the design of the overall sample size along with regional sample fractions. We also provide the theoretical footage for assessing consistency probability using Method 1 of \citet{japan_criteria}, based on the empirical shrinkage estimator. The latter is then used to determine the regional sample size of interest. We elaborate on the applications to common continuous, binary, and survival endpoints in detail. Simulation studies show that the proposed method retains the consistency probability at the desired level. We illustrate the application using a real cardiovascular outcome trial in diabetes. An R package is provided for implementation.

\end{abstract}


\newpage 
\section{Introduction}\label{sec:intro}

Under \cite{ICHE17}, multi-regional clinical trials (MRCTs) have been widely adopted in drug development as an efficient way to expedite the process and enable rapid availability of drugs to patients worldwide. Designs of MRCT and regional sample size determination have focused on continuous or binary endpoints under the fixed effects approach, which assumes a common treatment effect across regions \citep{ikeda2010sample, ko2010sample,chen_decision_2012,homma2024cautionary,qing2025twoMRCT}. Others have adopted a more flexible approach that allows the treatment effect to vary across regions (i.e., region-specific effect)  \citep{chen_assessing_2010,quan2010sample,tanaka2012qualitative, li2024regional}.

Yet, in many situations such as oncology and cardiovascular trials with long duration, regional treatment heterogeneity arises due to ethnicity reason or regional covariates (e.g., sex or age) shift \citep{Cappellini:2020}. The LEADER (Liraglutide Effect and Action in Diabetes) trial was a randomized, double-blind, placebo-controlled MRCT to determine the long-term effect of liraglutide on cardiovascular events in subjects with type 2 diabetes and high cardiovascular risk \citep{marso2016liraglutide}. A total of 9,340 patients were randomized 1:1 to liraglutide or placebo in four regions, namely, Europe (EU), North America (NA), Asia, and the rest of the world (ROW) with sample size fractions of 35\%, 30\%, 8\%, and 27\%, respectively. The primary endpoint was a survival endpoint with composite outcomes (i.e., time to the first occurrence of death from cardiovascular causes, nonfatal myocardial infarction, or nonfatal stroke). The overall hazard ratio was 0.87 (95\% CI: 0.78, 0.97), while the regional hazard ratios were 0.82, 1.01, 0.62, 0.83 for the four regions respectively. The result of North America differed greatly from others, despite its large contribution in participation. On the other hand, the Asia population exhibited a more pronounced treatment effect. 
All these indicated a high degree of heterogeneity across regions. The fixed effects approach becomes inappropriate. 

Inspired by meta-analysis, the random effects model was proposed for MRCT design, where regional effects are considered random variables with the mean representing the overall effect. \citet{chen_design_2012} proposed using a weighted average of regional sample means to estimate the overall treatment effect and determined the overall sample size by inverting a t-test. \citet{wu2014sample} proposed modeling the heterogeneous variability by imposing a gamma component in the regional variance and obtaining the maximum likelihood estimators of the parameters through the EM algorithm. The overall sample size is determined by inverting a z-test. 
Under the same model of \citet{chen_design_2012}, \citet{quan2013empirical} proposed using the Bayesian shrinkage estimator to estimate the regional effect. \citet{quan2014multi} further recommended the James--Stein type estimator for the regional effect to protect against over-shrinkage. Along the same lines, \citet{adall2021bayes} modified the empirical Bayesian shrinkage estimator to allow heterogeneous variances across regions and proposed a truncated version that mitigates risk in the presence of extreme values of regional treatment effects. 
All these methods adopt the criterion (Method 1) from the ``Basic Concepts on Global Clinical Trials'' guidance by \citet{japan_criteria} to assess the regional consistency of interest provided the overall significance and uses it to determine regional sample size 
However, the explicit expression of the assurance probability of the consistency based on the empirical shrinkage estimator has not been provided. Nor have its statistical properties been fully studied. Moreover, the application of the model and regional sample size calculation to the case of survival endpoints without proportional hazards is not seen. 

In this paper, we first review the inferential methods under the random effects model and then provide a theoretical footage of consistency probability and regional sample size calculation. The lower bounds of the consistency probability under some special cases are obtained. We apply the proposed method to the continuous, binary, and survival endpoints. The use of restricted mean survival time (RMST) for survival endpoint under non-proportional hazards is elaborated. These constitute the main contents of Section~\ref{sec: method}.  We demonstrate the performance of the design by simulation in Section~\ref{sec:simulation} and illustrate the application to a real example in Section~\ref{sec: realexample}. We conclude the paper with a discussion in Section~\ref{sec:disc}. 

\section{Method} \label{sec: method}

\subsection{Inferences under random effects model}\label{sec: emp shrinkage est}

\subsubsection{Estimation}\label{sec:estimation}
Consider a two-arm trial with subscript $k$ taking 0 and 1 to represent the control group and treatment group, respectively. Let $n_k$ denote the size of independent samples from group $k$ and $n=n_0+n_1$ denote the total sample size of the trial.

Suppose $R\ (\ge 2)$ regions participate in the trial. We use the subscript or superscript (depending on convenience) $r$ to represent statistics or quantities for region $r$.
Partition $n$ by regions as $n=n^{(1)}+\cdots+n^{(R)}$. Further partition regional sample size by group as $n^{(r)}=n^{(r)}_1+n^{(r)}_0$. Assume the randomization ratios between groups are fixed across regions, i.e., $\ell=n^{(r)}_1/n^{(r)}_0$ for all $r$. 
Let $f_r=n^{(r)}/n$ be the fraction of regional sample size with respect to the overall sample size.   

For region $r$, let $D_r$ denote the regional treatment effect. Let $\widehat{D}_r$ be a sensible estimator of $D_r$ and the distribution of $\widehat{D}_r$ (conditional on $D_r$) is approximately $N(D_r,\sigma^{2(r)})$. The variance $\sigma^{2(r)}$ is connected to the regional sample size fraction through 
\begin{equation}\label{eq: Omega}
\sigma^{2(r)}=\frac{\Omega_r}{n_0f_r},  
\end{equation}
where $\Omega_r$ depends on the variance of the responses and randomization ratio $\ell$, and is elaborated in Section~\ref{sec:application}.  

To accommodate variations of the  treatment effect across regions, the random effects model \citep{quan2013empirical} imposes a prior of $D_r$ as $D_r\sim N(\delta,\tau^2)$, where $\delta$ is the overall treatment effect and $\tau^2$ is the hyperparameter that governs the cross-region variation. 
Then, the unconditional distribution of $\widehat{D}_r$ is $N(\delta,\tau^2+\sigma^{2(r)})$.  
Let $w_r = (\tau^2+\sigma^{2(r)})^{-1}$ and $w = \sum_{r=1}^{R}w_r$. The overall treatment effect is estimated by
\begin{equation} \label{eq: WLSE}
     \widetilde{D} = \sum_{r=1}^{R}w_r\widehat{D}_r /w, 
\end{equation}
which is approximately distributed as $N(\delta, w^{-1})$ \citep{whitehead2002meta}. Note that $w^{-1}$ is smaller than the variance of the naive estimate $\widehat{D}=\sum_{r=1}^{R}f_r\widehat{D}_r $ under the random effects model. 

In practice, suppose one collects information on regional treatment $D_r$ from historical trials or literature. One naive approach is to estimate the hyperparameters $\delta$ and $\tau^2$ by the mean and variance of $D_r$s. An alternative estimator of $\tau^2$ is the moment estimator \citep{dersimonian1986meta} given by 
\begin{equation}\label{eq:tauvhat}
\textrm{max}\left\{0, 
\frac{\sum_rw_r(\widehat{D}_r-\widetilde{D})^2-(R-1)} {w-w^{-1}\sum_{r}w_r^{2}}\right\} 
\end{equation} 
with $\tau^2=0$ in $w_r$ defined earlier. However, caution needs to be taken as this moment estimator tends to underestimate $\tau^2$ especially when the heterogeneity variance is moderate or large \citep{sidik2007comparison}.  

At last, let $\widehat{\sigma}^{2(r)}$ and $\widehat{\tau}^2$ denote the estimates of $\sigma^{2(r)}$ and $\tau^2$ from the regional sample. Denote the plug-in estimators of $w_r$ and $w$ by $\widehat{w}_r=\widehat{\tau}^2+\widehat{\sigma}^{2(r)}$ and $\widehat{w}=\sum_r\widehat{w}_r$, respectively.

\subsubsection{Hypothesis testing and sample size calculation}\label{sec: sample-size-n}

Consider testing the hypothesis $H_0: \delta=0$ vs $H_1: \delta>0$ (assuming that larger value of $\delta$ is preferred). Let $T=\widetilde{D}/\widehat{\textrm{sd}}(\widetilde{D})$ be the test statistic, where $\widetilde{D}$ is obtained from (\ref{eq: WLSE}) with the plug-in estimates of $\widehat{w}_r$ and $\widehat{w}$, and $\widehat{\textrm{sd}}(\widetilde{D})=\widehat{w}^{-1/2}$. By large sample theory, we reject the null hypothesis when $T>z_{1-\alpha}$ with size $\alpha$, where $z_q$ is the $q$th quantile of standard normal distribution.  

For a power $1-\beta$ test, \citet{quan2013empirical} proposed to calculate the sample size $n_0$ by solving the equation
\begin{equation}\label{eq:nmrct}
\var^{-1}(\widetilde{D})=w=\sum_r\left(\tau^2+\frac{\Omega_r}{n_0f_r}\right)^{-1} =\frac{(z_{1-\alpha}+z_{1-\beta})^2}{\delta^2}.
\end{equation}
The overall sample size is computed numerically through (\ref{eq:nmrct}). We provide the R program in the Web Appendix. As seen in Table~\ref{tab:ss-sensitive-omega1}, $n_0$ is sensitive to the parameters of $\tau$ and $\delta$.   
Meanwhile, the specifications of $\tau$, $\delta$ and $\Omega_r$ are not arbitrary, as shown in later sections. In the design stage, one needs to elicit them from available information as accurate as possible.  

When the variance component $\Omega_r$s are the same, say $\Omega$, and 
the regional sample sizes are equally allocated (i.e., $f_r=R^{-1}$), the minimum overall sample sizes for the two groups are attained as 
\begin{equation} \label{eq:SShomo}
n_0=\frac{R\Omega(z_{1-\alpha}+z_{1-\beta})^2}{R\delta^2-\tau^2(z_{1-\alpha}+z_{1-\beta})^2},\quad n_1=\ell n_0.
\end{equation}
Note the positive denominator of (\ref{eq:SShomo}) intrinsically implies  $\tau/\delta<\sqrt{R}/(z_{1-\alpha}+z_{1-\beta})$. It is clear that $n_0$ decreases in $R$ as noted in \citet{quan2013empirical}. However, such homogeneity can hardly hold for the binary and survival endpoints.  

\subsection{Consistency assessment and regional sample size calculation} \label{sec: consistency}

\subsubsection{Regional effect estimator}
Consider the consistency assessment of the regional treatment effect with respect to the overall effect. Given the conditional distribution of $\widehat{D}_r$ and the prior distribution of $D_r$, the posterior distribution of $D_r$ is $N(\delta_r^*,\sigma^{2*(r)})$, where
\begin{equation}\label{eq: post-param}
\delta_r^*= \frac{\tau^{2}}{\tau^{2}+\sigma^{2(r)}}\widehat{D}_r
    + \frac{\sigma^{2(r)}}{\tau^{2}+\sigma^{2(r)}}\delta, \quad
\sigma^{2*(r)} =
    \frac{1}{\tau^{-2}+\sigma^{-2(r)}}.
\end{equation}
The regional treatment effect $D_r$ can be estimated by the empirical Bayesian shrinkage estimator by replacing $\delta$ with $\widetilde{D}$ in (\ref{eq: WLSE}) as
\begin{equation}\label{eq: shrinkage-est}
\widetilde{D}_r = \frac{\tau^{2}}{\tau^{2}+\sigma^{2(r)}}\widehat{D}_r
    + \frac{\sigma^{2(r)}}{\tau^{2}+\sigma^{2(r)}}\widetilde{D}.
\end{equation}
We have 
\begin{equation}\label{eq:cov-Dr-tilde}
\var(\widetilde{D}_r)=w_r\tau^4
    +\frac{\sigma^{2(r)}(2\tau^2+\sigma^{2(r)})}{w(\tau^2+\sigma^{2(r)})^2},\quad
\cov(\widetilde{D}_r,\widetilde{D})=\var(\widetilde{D})=w^{-1}.
\end{equation}
It can be verified that $w^{-1}=\var(\widetilde{D})\le \var(\widetilde{D}_r)\le \var(\widehat{D}_r)=\sigma^{2(r)}+\tau^2$. 
When $\tau=0$, i.e., no variation in mean value across regions, $\widetilde{D}_r$ coincides with $\widetilde{D}$, rather than the regional estimator $\widehat{D}_r$. In other word, $\widetilde{D}_r$ shrinks toward the overall effect instead of $\widehat{D}_r$. As noted by \citet{quan2013empirical}, the treatment effect for region $r$ may not be best described by $\widehat{D}_r$ which ignores data from other regions particularly when there is a significant overall treatment effect. By utilizing information from other regions, the empirical shrinkage estimator is a more precise estimator of regional treatment effect compared to the regional sample mean estimator.  

Let $\rho_r=\var(\widetilde{D})/\var(\widetilde{D}_r)$, which is in $(0,1]$. 
The joint distribution of $(\widetilde{D}_r,\widetilde{D})$ is approximately $N(\vmu, \Sigma)$ where
\begin{equation}\label{eq:mu-sigma}
\vmu=\delta(1,1)^\top,\quad
\Sigma = w^{-1}
\begin{pmatrix}
    \rho_r^{-1} & 1 \\
    1 & 1
\end{pmatrix}
,\quad
\rho_r^{-1}=ww_r\tau^4 +\frac{\sigma^{2(r)}(2\tau^2+\sigma^{2(r)})}{(\tau^2+\sigma^{2(r)})^2}.
\end{equation}

To facilitate subsequent derivation, let $h_r=\tau^2/\sigma^{2(r)}$, which relates the regional fraction through (\ref{eq: Omega}). Then, we can express 
\begin{equation}\label{eq: expression-in-hr}
w=\frac{1}{\tau^2}\sum_{j=1}^R\frac{h_j}{h_j+1},\quad \rho_r^{-1}=1+\frac{h_r}{h_r+1}\sum_{j \ne r}^R\frac{h_j}{h_j+1}.
\end{equation}  
Note that both $w$ and $\rho_r^{-1}$ involve $n_0$ and $f_r$s at the same time. 

\subsubsection{Consistency probability and regional sample size determination} \label{sec:cp}

The criterion in Method 1 of \citet{japan_criteria} requires that the consistency probability of the fraction of regional treatment effect of interest with respect to the overall effect be no less than $\pi \ (\ge 0.5)$ given the significance of the overall treatment effect is no less than the prespecified threshold $1-\gamma\ (\ge 0.8)$.
Using the shrinkage estimator of the treatment effect of region $r$, the criterion amounts to
\begin{equation}\label{eq:criterion}
    \prob\left(\widetilde{D}_{r} \ge \pi \widetilde{D} \vert T>z_{1-\alpha}\right) \ge  1-\gamma.
\end{equation}
Note that \citet{quan2013empirical} computed the unconditional probability of $\widetilde{D}_{r} \ge \pi \widetilde{D}$.

\begin{proposition} \label{prop:criterion-approxv2}
Under the alternative hypothesis with $\delta >0$, the consistency probability specified by the LHS of (\ref{eq:criterion}) is approximately
\begin{equation}\label{eq:criterion-approxv2}
\frac{1}{1-\beta}\int_{u>z_{\beta}}\Phi\left(\frac{(1-\pi) \left(u+z_{1-\alpha}+z_{1-\beta}\right) }{\sqrt{\rho_r^{-1}-1}}\right) d\Phi(u)
\end{equation}
where $\rho_r^{-1}$ is defined in (\ref{eq: expression-in-hr}).
\end{proposition} 
Recall that by (\ref{eq:nmrct}) the overall sample size is determined given $f_1,\ldots,f_R$. This is different from the determination of $f_r$ in the fixed effects model, where the overall sample size is free of regional fractions \citep{ko2010sample,quan2010sample,qing2025twoMRCT}. The monotonicity of CP in (\ref{eq:criterion-approxv2}) with respect to $f_r$ is complicated, as the term $\rho_r^{-1}$ involves both $n_0$ and $f_1, \ldots, f_R$ through (\ref{eq: expression-in-hr}). By (\ref{eq: Omega}) and (\ref{eq: shrinkage-est}), the degree by which the regional effect estimator $\widetilde{D}_r$ shrinks towards the overall estimator $\widetilde{D}$ decreases in $n_0f_r$.  This may lead to some counter-intuitive results about the relationship between CP and $f_r$ as shown in the following example.

\begin{example}\label{ex:foequal}
Suppose $\Omega_r=2$ for all $r$. Set $\delta=0.25$, $\tau=0.1$, $\ell=1$, $\pi=0.5$, $\alpha=0.025$ and $\beta=0.1$.
Figure~\ref{fig:both-f1} shows the approximated CP of region $r$ by (\ref{eq:criterion-approxv2}) as a function of $f_r$ when samples are evenly allocated among other regions (i.e., $f_j=(1-f_r)/(R-1)$ for $j\ne r$) for $R=3,4$. The CPs (right panel) are seen to decrease in $f_r$ until around $0.8$, then nearly plateau afterwards. This is consistent with the fact that the parabola-shaped $\sqrt{\rho_r^{-1}-1}$ (left panel) has its maximum occurring around $f_r =0.8$. In addition, we see CP is no less than 96\% for $f_r \in (0,0.8)$, indicating that CP is high and stable across various configurations of regional fractions $(f_1,\dots,f_R)$. The detailed values of overall sample size, regional sample size and CP are provided in Table~\ref{tab:CP-f1desc}.  
\end{example}
 
\begin{corollary}\label{co:cplowerbound}
The lower bound of the consistency probability in (\ref{eq:criterion-approxv2}) is attained at
\begin{equation}\label{eq:cplowerbound-v2}
  \frac{1}{1-\beta}\int_{u>z_{\beta}}\Phi\left(\frac{2\delta^2(1-\pi)(u+z_{1-\alpha}+z_{1-\beta})}{\tau^2(z_{1-\alpha}+z_{1-\beta})^2} \right) d\Phi(u).
\end{equation}
when $\frac{h_r}{h_r+1}=\frac{\tau^2}{2\delta^2}(z_{1-\alpha}+z_{1-\beta})^2<1$ for the region of interest. 
\end{corollary}  

The lower bound decreases in $\tau/\delta$ as expected. The attainability is also restricted by the inequality condition, i.e., ${\tau}/{\delta}< {\sqrt{2}}/{(z_{1-\alpha}+z_{1-\beta})}$. For example, when $(\alpha,\beta)$ takes on $(0.025,0.1)$, $(0.025, 0.2)$, $(0.05,0.1)$ and $(0.05,0.2)$, ${\tau}/{\delta}$ needs to be less than 0.44, 0.50, 0.48, and 0.57, respectively. When the inequality condition in Corollary \ref{co:cplowerbound} fails to hold, the lower bound (\ref{eq:cplowerbound-v2}) is not attainable. Nevertheless, we can show the CP is greater than  
\begin{equation}\label{eq:cplowerbound-v3}
\frac{1}{1-\beta}\int_{u>z_{\beta}}\Phi\left(\frac{(1-\pi) \left(u+z_{1-\alpha}+z_{1-\beta}\right) }{\sqrt{\frac{\tau^2}{\delta^2}(z_{1-\alpha}+z_{1-\beta})^2-1}}\right) d\Phi(u).
\end{equation} 

\begin{example}\label{ex:cplbd}
Table~\ref{tab:CPlowerbound-prop2} presents the lower bounds of the approximate CP of (\ref{eq:cplowerbound-v2}) under the combinations of $\alpha\in \{0.025,0.05\}$, $\beta\in \{0.1,0.2\}$ and $\tau/\delta\in\{0.4,0.6\}$ given $\pi=0.5$. It is seen that the lower bounds of CP are all greater than 80\% for the considered values of $\tau/\delta$, which represent an appropriate range for an MRCT. Further assume $\Omega_1=2$ in region 1 of interest
and $\ell=1$. Two (out of many) possible solutions of $(n_0, f_1)$ are provided.
\end{example}

\begin{corollary}\label{co:criterion-approxv2-fequal}
When samples are equally allocated to regions, i.e., $f_r=R^{-1}$ for all $r$, and the variances are homogeneous across the regions, i.e., $\Omega_r=\Omega$ for all $r$, the consistency probability in (\ref{eq:criterion-approxv2}) is simplified as 
    \begin{equation}\label{eq:criterion-approx-fequal}
    \frac{1}{1-\beta}\int_{u>z_{\beta}}\Phi\left(\frac{\delta^2R(1-\pi) \left(u+z_{1-\alpha}+z_{1-\beta}\right)}{\tau^2 \sqrt{R-1}(z_{1-\alpha}+z_{1-\beta})^2 }\right) d\Phi(u).
    \end{equation}
provided that $\tau/\delta<\sqrt{R}/(z_{1-\alpha}+z_{1-\beta})$.  
\end{corollary}
 
Table~\ref{tab:CP-fequal} presents the results of CP under combinations of $\alpha\in\{0.025, 0.05\}$, $\beta\in \{0.1, 0.2\}$, $R\in\{3, 4\}$ and $\tau/\delta\in\{0.4, 0.6\}$ with $\pi=0.5$. The CPs are all no less than 85\%.  

In summary, under the random effects model, the regional sample size/fraction and the total sample size are determined at the same time. One can use Proposition~\ref{prop:criterion-approxv2}, Corollaries~\ref{co:cplowerbound} and \ref{co:criterion-approxv2-fequal} to configure the regional fraction of interest to warrant the desired CP. 
We provide an R package that computes the overall sample size and consistency probability in Supplementary Materials and in GitHub (\texttt{https://github.com/kunhaiq/ssMRCT.git}).

The proposed method based on the superiority trial is readily extendable to the non-inferiority (NI) trial, where the hypothesis in Section~\ref{sec: sample-size-n} becomes $H_0: \delta=-M$ vs.\ $H_1: \delta>-M$, with $M>0$ being the prespecified NI margin. One just needs to i) modify the test statistic to be $T=(\widetilde{D}+M)/\widehat{\textrm{sd}}(\widetilde{D})$, ii) replace $\delta$ with $\delta+M$ in (\ref{eq:nmrct}), and iii) modify the consistency criterion as $\widetilde{D}_r+M \ge \pi (\widetilde{D}+M)$. The design procedure remains the same as for the superiority trial. For consistency assessment, Proposition~\ref{prop:criterion-approxv2} and its corollaries hold after replacing $\delta$ with $\delta+M$.

\subsection{Application to various endpoints}\label{sec:application}

\subsubsection{Continuous and binary endpoints}\label{sec:appcts}

Let $Y^{(r)}_{k}$ denote the response of either continuous or binary type from group $k$ in region $r$. The regional treatment effect is $D_r=\E(Y^{(r)}_{1})-\E(Y^{(r)}_{0})$. For continuous response, the variance $\sigma^{2(r)}_k$ is constant. 
For binary response, the variance $\sigma^{2(r)}_k$ equals $\E(Y^{(r)}_k) \{1-\E(Y^{(r)}_k)\}$. Then, the quantity $\Omega_r$ in (\ref{eq: Omega}) associated with the variance of the regional treatment effect estimate is expressed as $\Omega_r = \ell^{-1}\sigma^{2(r)}_1+\sigma^{2(r)}_0$.

\subsubsection{Survival endpoint under proportional hazard assumption}\label{sec:appsurv}

Suppose the response is a survival endpoint satisfying the proportional hazard (PH) assumption. 
The regional treatment effect $D_r$ is the negative log hazard ratio (HR) between treatment and control. Let $N_r$ denote the expected number of events in region $r$. 
Let $\widehat{D}_r$ denote the estimate from the Cox model. 
By \citet{schoenfeld1981asymptotic}, the conditional distribution of $\widehat{D}_r$ is approximately $N(D_r, \sigma^{2(r)})$, where $\sigma^{2(r)}=(\ell+1)^2/\{\ell N_r\}$. 

To express $\sigma^{2(r)}$ in the form of $\Omega_r$ and the regional sample size as in (\ref{eq: Omega}), additional assumptions are needed on the survival distribution, censoring, and trial design. For simplicity, consider a trial with a fixed study duration in which all subjects are followed for the same amount of time, say $L$ time units. Let $T_k^{(r)}$ denote the survival time of group $k$ in region $r$. Assume that $T_k^{(r)}$ follows the exponential distribution with rate parameter $\lambda_k^{(r)}$ and that administrative censoring (at the end of the study) is adopted. Then, we have $\Omega_r =(\ell+1)^2/\{\ell (P_0^{(r)} + \ell P_1^{(r)})\}$, where $P_k^{(r)}=1-e^{-\lambda_k^{(r)}L}$ is the probability of having an event during the study for subject of group $k$ in region $r$.

\subsubsection{Survival endpoint under non-proportional hazard assumption}\label{sec:appsurv-nPH}

When the proportional hazard assumption is violated for the survival endpoint, such as by a delayed effect of immunotherapy in an oncology trial or by low event rates in a cardiovascular trial, the interpretation of HR becomes inappropriate and the Cox PH test or log-rank test loses power \citep{zhao2012utilizing, uno2014moving}. In this case, RMST-based between group comparison is recommended as a meaningful non-parametric alternative to HR. 
Intuitively, RMST is the expected value of survival time restricted to a prespecified time point $\eta$ and equals the area under the survival curve up to $\eta$. Specifically, the RMST of group $k$ is $\mu_k = \E\{\min(T_k,\eta)\}=\int_{0}^{\eta} S_k(t)dt$, where $T_k$ is the survival time of group $k$ with associated survival function $S_k(t)$. In practice, $\eta$ is usually chosen as a prespecified constant allowing a straightforward interpretation of $\mu_k$, such as the end-of-study time or the largest observed time \citep{tian2020empirical}.
Naturally, we use the difference of the RMSTs, $D=\mu_1-\mu_0$, to measure the treatment effect.

Let $\widehat{\mu}_k=\int_{0}^{\eta} \widehat{S}_k(t)dt$ be a sensible estimator of $\mu_k$, where $\widehat{S}_k(\cdot)$ is the Kaplan-Meier estimator. \citet{pepe1989weighted} showed that $\sqrt{n_k}(\widehat{\mu}_k - \mu_k)$ is asymptotically distributed as $N(0,\sigma_k^2)$, where $\sigma_k^2$ is given in Appendix \ref{sec:techDetailSur}.
Let $\widehat{D}=\widehat{\mu}_1-\widehat{\mu}_0$. Then, conditional on $D$, $\widehat{D}$ is approximately $N(D, \sigma^{2})$, where $\sigma^{2}=n_1^{-1}\sigma^2_{1} + n_0^{-1}\sigma^2_{0}$. 
Applying these results to region $r$, we get $D_r$, $\widehat{D}_r$, and $\sigma^{2(r)}$ accordingly, and $\Omega_r = \ell^{-1}\sigma_1^{2(r)}+\sigma_0^{2(r)}$ as in the continuous case.

\section{Simulation}\label{sec:simulation}

\subsection{General setup and procedure} \label{sec:setup}
Consider $R=3$ and $4$. Set balanced randomization between groups in all regions, i.e., $\ell=1$. Also, set $\alpha=0.025$, $\beta=0.2$, and $\pi=0.5$. Without loss of generality, suppose region 1 is the region of interest to be assessed for consistency. 

We summarize the procedures for the simulation as follows.
\begin{quote}
\begin{itemize}
  \item[Step 1:] [Benchmark] Specify the regional treatment effects and the regional variances in $\{(D_r, \Omega_r): r=1,\ldots,R\}$ and the regional fractions $f_1,\ldots,f_R$. Use the naive approach in Section~\ref{sec:estimation} to compute $\delta$ and $\tau$. Compute the overall sample size $n_0$ by (\ref{eq:nmrct}) and CP of region 1 by (\ref{eq:criterion-approxv2}).  These will be used as benchmarks for reference.  

  \item[Step 2:] [Design] Mimic the actual design to obtain sensible estimates of $(D_r, \Omega_r)$ based on samples of moderate to large size from each region specified in Step~1. Denote them by $(D_r^{\rm D}, \Omega_r^{\rm D})$. Repeat the same approach as in Step 1 to compute $\delta$ and $\tau$, denoted by $\delta^{\rm D}$ and $\tau^{\rm D}$, and subsequently the overall sample size and CP, denoted by $n_0^{\rm D}$ and $\textrm{CP}^{\rm D}$.  
      
  \item[Step 3:] [Verification] Based on $\delta^{\rm D}$, $\tau^{\rm D}$, and $n_0^{\rm D}$ obtained in Step 2 and $f_1 \ldots, f_R$ specified in Step 1, generate regional treatment effects and regional data. (The regional distribution depends on the type of endpoint and is elaborated in Section \ref{sec:setupSpec}.) Perform the test and consistency evaluation $m^{\rm V}$ times to compute the empirical power and CP. In addition, compute the absolute deviation of the empirical power from the nominal power, denoted by ${\rm dev}(\beta)=|{\rm empirical\ power}-(1-\beta)|$, and the absolute deviation of the empirical CP from the designed value from Step 2, denoted by ${\rm dev(CP)}=|{\rm empirical\ CP}-\textrm{CP}^{\rm D}|$.   
\end{itemize}
\end{quote}
Repeat Steps 2 and 3 for $m^{\rm D}$ times and compute i) the median of $n_0^{\rm D}$, denoted by $\widetilde{n}_0^{\rm D}$, and the average of $\textrm{CP}^{\rm D}$, denoted by $\overline{\textrm{CP}}^{\rm D}$, and ii) the averages of ${\rm dev}(\beta)$ and ${\rm dev(CP)}$, denoted by $\overline{\rm dev}(\beta)$ and $\overline{\rm dev}({\rm CP})$. The latter two are used as metrics to examine the performance of the proposed design.  

Since $n_0$ is sensitive to the values of $\tau$ and $\delta$, $n_0^{\rm D}$ can deviate remarkably from the benchmark value depending on the precision of the estimated parameters compared to their benchmark values. Therefore, we also compute the median of the absolute deviation (MAD) of $n_0^{\rm D}$ associated with $\widetilde{n}_0^{\rm D}$ to examine such deviation.  

Throughout, set the training sample size used in Step 2 to be 1,000 per group per region, and set the numbers of replications $m^{\rm D}$ and $m^{\rm V}$ both to be 1,000.

\subsection{Specific setup for different endpoints}\label{sec:setupSpec}
      
\subsubsection{Continuous and binary endpoints}\label{sec:ctsOutcome}

Specify $D_r$ and $f_r$ in the first three columns of Table~\ref{table:sim-result-ss1000-normal-binary}.
For the continuous endpoint, set $\Omega_r=2$ by letting $\sigma_0^{2(r)}=\sigma_1^{2(r)}=1$ for all $r$. Set $Y^{(r)}_{0}\sim N(0,\sigma_0^{2(r)})$ and thereby $Y^{(r)}_{1}\sim N(D_r^{\rm V},\sigma_1^{2(r)})$.
For the binary endpoint, set $\E(Y_0^{(r)})=0.3$ for all $r$, and thus the distribution of $Y_1^{(r)}$ is Bernoulli with parameter $0.3+D_r^{\rm V}$.

\subsubsection{Survival endpoint with PH assumption}\label{sec:surOutcome-PH}
Besides the setups in Section~\ref{sec:appsurv}, assume that $T_0^{(r)}$ follows the exponential distribution with rate parameter $\lambda_0^{(r)}$. Set $\lambda_0^{(r)}=0.05$ for all $r$. Specify the regional treatment effect in terms of $\textrm{HR}_r$ and $f_r$ in the first three columns of Table~\ref{table:sim-result-ss1000-PH-dev-round}. Then, the distribution of $T_1^{(r)}$ is the exponential distribution with parameter $\lambda_1^{(r)}=\lambda_0^{(r)}\times\textrm{HR}_r$. Set the study duration $L$ to be 36 time units. In the verification step, the parameter of the distribution of $T_1^{(r)}$ is obtained by $\lambda_1^{(r)}$=$\lambda_0^{(r)} e^{-D_r^{\rm V}}$.   

\subsubsection{Survival endpoint with non-PH assumption}\label{sec:surOutcome-nonPH}

Consider three scenarios of two survival curves that represent early effect, late effect and crossing effect, respectively. 
For the first two scenarios (early and late effect), assume $T_{k}^{(r)}$ follows the piecewise exponential distribution with the hazard given as $\lambda_k^{(r)}I(0<t\le \psi )+\gamma_k^{(r)}I(t> \psi)$, where $\psi$ is the change point. Set $\lambda_0^{(r)}>\lambda_1^{(r)}$ and $\gamma_0^{(r)}=\gamma_1^{(r)}$ in the early effect scenario, and $\lambda_0^{(r)}=\lambda_1^{(r)}$ and $\gamma_0^{(r)}>\gamma_1^{(r)}$ in the late effect scenario. For the crossing scenario, assume $T_{k}^{(r)}$ follows the Weibull distribution with shape parameter $\nu_k^{(r)}$ and scale parameter $\theta_k^{(r)}$, denoted by Weibull$(\nu_k^{(r)}, \theta_k^{(r)})$. Set the distributions in such a way that in each region the survival curve of the control group is above that of the treatment group at the beginning and then crosses as time progresses. We defer all the specification details in Table~\ref{table:result-3-sce1} and illustrate these scenarios in Figures~\ref{fig:survnph-early-pw}--\ref{fig:scenario9}. The resulting $D_r$ and $\sigma^{2(r)}$ (through (\ref{eq: Omega}) with $\Omega_r= \sigma_0^{2(r)}+\ell^{-1}\sigma_1^{2(r)}$) are provided in Table~\ref{table:result-3-sce1} too. 

In all three scenarios, set $\eta=80$ and assume $C_{k}^{(r)}$ follows the uniform distribution on $(0,3\eta)$ for all $r$ and $k$, to reflect a drop-out rate of 5\% per 12 time units.   

For the verification step, the distribution of $T_{1}^{(r)}$ for the early and late effect scenario is obtained by solving for $\gamma_1^{(r)}$ such that $\mu_1^{(r)}-\mu_0^{(r)}=D_r^{\rm V}$, given $\eta$ and $\{\lambda_0^{(r)}, \gamma_0^{(r)}, \lambda_1^{(r)} \}$.  
For the crossing scenario, the distribution of $T_{1}^{(r)}$ is obtained by solving for $\nu_1^{(r)}$ such that $\mu_1^{(r)}-\mu_0^{(r)}=D_r^{\rm V}$, given $\eta$ and $\{\nu_0^{(r)}, \theta_0^{(r)}, \theta_1^{(r)}\}$.

\subsection{Results}

In Tables~\ref{table:sim-result-ss1000-normal-binary}--\ref{table:sim-result-ss1000-nPH-dev}, we first provide the benchmark values of $n_0$ and the CP of region~1 based on the true parameters. 
Second, we report two average values of $n_0^{\rm D}$ (through the median) and $\textrm{CP}^{\rm D}$ by design with parameters estimated from the simulated training data. The last two columns of the tables report two average values of the absolute deviation of the empirical power and the absolute deviation of the empirical CP.   

It is seen that the median of $n_0^{\rm D}$ and the average CP by design are nearly identical to their respective benchmark values, indicating the convergence of the solution of (\ref{eq:nmrct}). 
The wide range (from 13 to 419) of the MAD of $n_0^{\rm D}$ depends on the value of $n_0$ proportionally, which again indicates the sensitivity of $n_0$ with respect to the design parameters.  
 
As of our greatest concern, the average absolute deviations of the empirical power and empirical CP (from the designed values) vary in 1.2\%--6.8\% and 1.3\%--4.8\%, respectively, across all considered scenarios and endpoints, regardless of whether the configuration of regional fractions is uniform or varying across regions. These results show the proposed design is effective in retaining the power and CP at the desired levels.  
Note that ${\rm dev(CP)}$ depends on the ratio of $\tau/\delta$ (i.e., the standardized between-region variability) as expected. For instance, in the scenarios of the binary endpoint and the survival endpoint with the PH assumption, the two cases under three regions with $\tau/\delta$ of 0.5 yield $\overline{\rm dev}(\rm CP)$ of 2.7\% and 1.7\%, respectively. In contrast, the two cases under four regions with $\tau/\delta$ of 0.6 for these two endpoints lead to $\overline{\rm dev}(\rm CP)$ of about 4.0\% and 2.9\%, respectively.  

\section{Real example} \label{sec: realexample}

Return to the motivation example of the LEADER trial. The trial used a closed testing procedure to test non-inferiority of liraglutide versus placebo with margin of HR=1.3 (or equivalent $M=\log(1.3)=0.26$) first and then test for superiority if NI is significant. (No adjustment of the significance level is required.)  
Overall, the primary outcome occurred in significantly fewer patients in the liraglutide group ($608/4668$ [13.0\%]) compared with placebo ($694/4672$ [14.9\%]). The HR (95\% CI) was 0.87 (0.78, 0.97) with a p-value less than 0.001 for noninferiority and a p-value of 0.01 for superiority. Subgroup analysis by region (Table~\ref{table:LEADER-results}) reveals that only Europe was significant at the nominal size of $0.025$ (two-sided).  

\begin{figure}[t]
    \centering
    \includegraphics[width=0.8\linewidth]{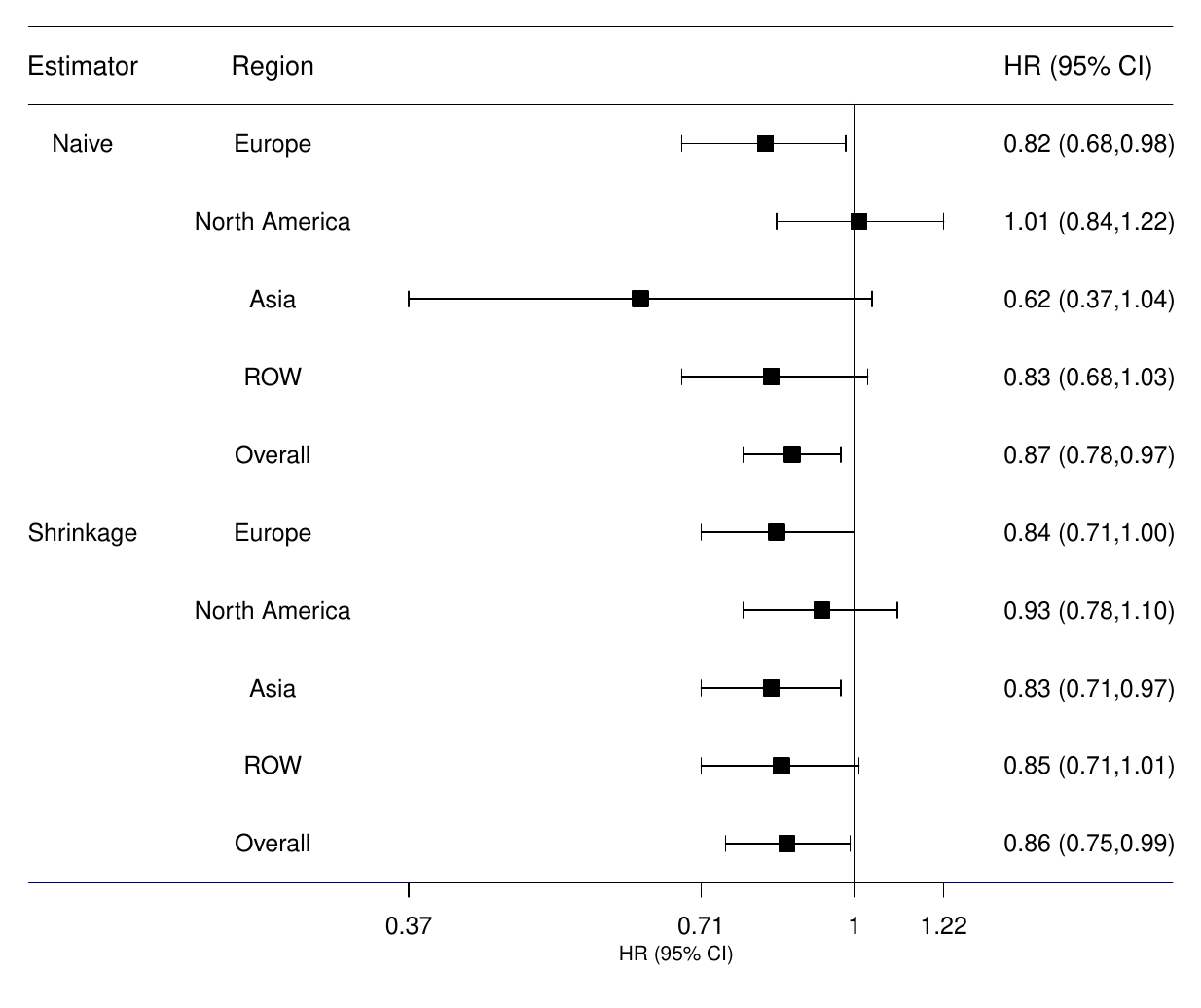}
    \caption{Comparison of regional HR (95\% CI) of the primary endpoint by region based on the naive estimator under the fixed effects model and the shrinkage estimator under the random effects model }
    \label{fig:LEADER}
\end{figure}

As pointed out in Section~\ref{sec:intro}, there appeared a remarkable heterogeneity in treatment across regions. Now, we adopt the random effects model to re-analyze the data. By the method described in Section~\ref{sec:appsurv} and replacing the expected number of events with the observed number of events in the LEADER trial, we obtain $\widehat{\sigma}^{2(r)}$ (given in Table~\ref{table:LEADER-results}), 
$\widehat{\tau}^2=0.0077$ by (\ref{eq:tauvhat}), and 
$\widetilde{D}=0.15$ with the estimated variance $\widehat{w}^{-1}=0.005$. 
We present the estimated overall HR (95\% CI) and regional HR in contrast with their counterparts under the fixed effects model in Figure~\ref{fig:LEADER}. It is seen that the point estimates of regional effects based on the shrinkage estimator were centered around the overall effect and were all less than one, with EU and Asia both significant at a size of 0.05 (two-sided). 

For consistency assessment, the regional treatment effects based on $\widetilde{D}_r$ (i.e., $-\log\widetilde{\textrm{HR}}_r$) are 0.17, 0.08, 0.18 and 0.17 for Europe, North America, Asia, and ROW, respectively. By criterion (\ref{eq:criterion}) with $\pi=0.5$, all regions are consistent in both NI test and superiority test.  In contrast, the estimated regional effects based on $\widehat{D}_r$ reported under the fixed effects model are 0.20, $-0.01$, 0.48, and 0.19 for Europe, North America, Asia, and ROW, respectively, and the estimated overall effect was 0.14. North America was clearly not consistent with the overall population in for superiority.  

Next, we use the published data to illustrate the application of the proposed method in study design for a NI trial with $H_0: \delta \le -M$ vs $H_1: \delta>-M$. Assume the survival time follows the exponential distribution with parameter $\lambda_k^{(r)}$ for group $k$ in region $r$.  
We still use the NI hypothesized values in LEADER with $\delta$=0 and $\lambda_k^{(r)}=0.018$ per year for both $k$ and all $r$ and only adopt the estimate $\widehat{\tau}^2= 0.0077$ from the published results for $\tau^2$. (It would be unfair and not appropriate to use the published regional results of HR to re-design the trial as the results demonstrated superiority.)
Consider fixed duration study design with $L$ being 3.8 and 4.5 years, corresponding to the median follow-up time and the maximum follow-up as reported. We further assume no dropout during the study as in \citet{quan2010sample}. Then, the estimates of $\Omega_r$ can be obtained as described in Section~\ref{sec:appsurv}. Consider three configurations of regional fractions $(f_1, \ldots, f_4)$ in Table~\ref{table:LEADER-results-design-NI-LEADERassump}. The first two in each panel are common choices used in simulation, and the third is the one reported in LEADER. Set the same two types of errors $\alpha=0.025$ (one-sided) and $\beta=0.1$ as in LEADER, and set $\pi=0.5$ and $\ell=1$. 

The results of the overall sample size and regional CP are provided in Table~\ref{table:LEADER-results-design-NI-LEADERassump}. It is seen that the overall sample sizes are larger than those of LEADER. Longer study duration implies less sample size as expected. The increment in sample size comparing to that of the fixed effect model is expected as the extra component of between-region variability is introduced \citep{quan2013empirical}. On the other hand, CPs under all considered scenarios are greater than 99\%. It is warranted by the assumption of the  standardized between region variability $\tau/(\delta+M)=0.33$ through (\ref{eq:cplowerbound-v2}).

\section{Discussion} \label{sec:disc}

In this paper, we provide mathematical rigor for the design of MRCTs, including regional consistency evaluation and regional sample size calculation under the random effects model. Various types of endpoints are considered. We provide an R package to facilitate implementation in practice. 
As the random effects model assumes heterogeneity of regional effects, it generally requires a larger overall sample size than that under the fixed effects model. With such a high price and the nature of the shrinkage regional estimator, high consistency probabilities of each region are warranted. We will compare the two model approaches more thoroughly in a separate report.  

When applying the design to survival endpoint, we have used simple follow-up rule and censoring mechanism. In practice, one needs to compute the design parameters according to the actual situation, such as i) participants with random dropoffs during the study, or ii) participants are censored once the number of total required events are reached, or both. 

The consistency criterion for the non-inferiority trial considered is one natural extension from the superiority trial. There are other ways to define consistency, such as regional effect greater than the NI margin or regional effect greater than the lower 95\% CI of the overall effect, which are worth further investigation.

\section*{Acknowledgement}
The research was supported by a fund of MSD R\&D (China) Co., LTD.


\begin{table}[t]
    \centering
    \caption{Sample size per group and CP of region 1 by design and actual deviations of power and CP under {\it normal endpoint} and {\it binary endpoint} (specified in Section~\ref{sec:ctsOutcome}), where $D_r$ and $f_r$ are given in the first three columns, $n_0$ and CP are the benchmark values, $\widetilde{n}_0^{\rm D}$ and  $\overline{\textrm{CP}}^{\rm D}$) are the averages of the designed values with estimated parameters. }
    \begin{tabular}{ccc ccc ccc}
    \hline
    $R$&  $(D_1,\ldots,D_R)$&  $(f_1,\ldots,f_R)$&
    $n_0$& $\textrm{CP}$& 
   $\widetilde{n}_0^{\rm D}$(MAD) & $\overline{\textrm{CP}}^{\rm D}$ 
    & $\overline{\rm dev}(\beta)$ & $\overline{\rm dev}({\rm CP})$ \\
    \hline
    \multicolumn{3}{c}{normal endpoint} \\
    3&  (0.6,0.4,0.2)& (1/3,1/3,1/3)&
    284& 0.94&
    284 (97)&0.94&0.023& 0.018\\
    
    & & (0.2,0.3,0.5)& 
    312& 0.95& 
    312 (87)& 0.95& 0.022& 0.015\\
   
    4& (0.8,0.6,0.4,0.2)& (1/4,1/4,1/4,1/4) & 134& 0.94& 
    134 (19)& 0.94& 0.025& 0.014\\

    & & (0.1,0.2,0.3,0.4)& 152& 0.97& 
    152 (24)& 0.96& 0.023& 0.013\\

    \\
    
    \multicolumn{3}{c}{binary endpoint} \\

    3&  (0.6,0.4,0.2)& (1/3,1/3,1/3)&
    56& 0.94&
    56 (13)&0.94&0.037& 0.027\\
    
    & & $(0.2,0.3,0.5)$& 
    60& 0.95& 
    60 (14)& 0.94& 0.038& 0.026\\
    
    4& (0.6,0.4,0.3,0.1)& (1/4,1/4,1/4,1/4)& 89& 0.88& 
    89 (20)& 0.89& 0.041& 0.039\\  

    &  & $(0.1,0.2,0.3,0.4)$& 102& 0.90& 
    101 (25)& 0.90& 0.039& 0.040\\  
    \hline    
    
    \end{tabular}
    \label{table:sim-result-ss1000-normal-binary}
\end{table}

\begin{table}[t]
    \centering
    \caption{Sample size per group and CP of region 1 by design and actual deviations of power and CP under {\it survival endpoint with PH assumption} (specified in Section~\ref{sec:surOutcome-PH}), where regional HRs and $f_r$ are given in the first three columns. The remaining columns are defined the same as Table~\ref{table:sim-result-ss1000-normal-binary}.}
    \begin{tabular}{*{9}{c}}
    \hline
    $R$&  $(\textrm{HR}_1,\ldots,\textrm{HR}_{\rm R})$&  $(f_1,\ldots,f_R)$&
    $n_0$& $\textrm{CP}$& 
    $\widetilde{n}_0^{\rm D}$ (MAD)& $\overline{\textrm{CP}}^{\rm D}$ 
    & $\overline{\rm dev}(\beta)$ & $\overline{\rm dev}({\rm CP})$ \\
    \hline
    3& (0.7,0.6,0.4)& (1/3,1/3,1/3)& 168& 0.95& 
    158 (47)& 0.95& 0.034& 0.017\\

    & & (0.2,0.3,0.5)& 183& 0.95& 
    178 (56)& 0.95& 0.034& 0.016\\
    
    4& (0.8,0.7,0.5,0.4)& (1/4,1/4,1/4,1/4)& 210& 0.90& 
    210 (56)& 0.90& 0.068& 0.029\\

    & & (0.1,0.2,0.3,0.4)& 244& 0.92& 
    243 (68)& 0.92& 0.067& 0.028 \\
    
    \hline    
    
    \end{tabular}
    \label{table:sim-result-ss1000-PH-dev-round}
\end{table}

\begin{table}[t]
    \centering
     \caption{  
     Sample size per group and CP of region 1 by design and actual deviations of power and CP under {\it survival endpoint with non-PH assumption} (specified in Section~\ref{sec:surOutcome-nonPH}), where $D_r$ and $f_r$ are given in the second column and Table~\ref{table:result-3-sce1}. The remaining columns are defined the same as Table~\ref{table:sim-result-ss1000-normal-binary}. }
    \begin{tabular}{*{8}{c}}
    \hline
    Scenario& $(f_1, \ldots, f_R)$&
    $n_0$& $\textrm{CP}$& 
    $\widetilde{n}_0^{\rm D}$(MAD)& $\overline{\textrm{CP}}^{\rm D}$ 
    & $\overline{\rm dev}(\beta)$ & $\overline{\rm dev}({\rm CP})$ \\
    \hline

    1 (early effect)& 
    (1/4,1/4,1/4,1/4)& 572& 0.90& 
    528 (191)& 0.90& 0.013& 0.042\\
                    & 
    (0.4,0.3,0.2,0.1)& 639& 0.89& 
    601 (220)& 0.89& 0.015& 0.042\\
    
    2 (late effect)& 
    (1/4,1/4,1/4,1/4)&220& 0.93& 
    229 (58)& 0.92& 0.033& 0.045\\
                   &
    (0.4,0.3,0.2,0.1)& 249& 0.92& 
    259 (66)& 0.91& 0.038& 0.048\\

    3 (crossing)& (1/4,1/4,1/4,1/4)&856& 0.87& 
    681 (291)& 0.89& 0.029& 0.022\\
    & 
    (0.4,0.3,0.2,0.1)& 1066& 0.86& 
    846 (419)& 0.88& 0.020& 0.024\\

    \hline    
    
    \end{tabular}
    \label{table:sim-result-ss1000-nPH-dev}
\end{table}

\begin{table}[t]
    \centering
    \caption{Overall sample size per group ($n_0^\textrm{D}$) and CPs (CP$^\textrm{D}_r$) under various configurations of regional fractions and assumption of Section~\ref{sec:appsurv} with other parameters estimated from the LEADER trial ($\alpha=0.025$, $\beta=0.1$, $\pi=0.5$, and $\ell=1$)}
    \begin{tabular}{*{4}{c}}
    \hline
         $L$ (years) & $(f_1\ldots f_4)$& $n_0^\textrm{D}$& $\textrm{CP}^\textrm{D}_1$, $\textrm{CP}^\textrm{D}_2$, $\textrm{CP}^\textrm{D}_3$, $\textrm{CP}^\textrm{D}_4$\\
    \hline  
        3.8& $(1/4, 1/4, 1/4, 1/4)$& 6540   & all 99.7\%\\
        &$(0.1, 0.2, 0.3, 0.4)$& 6974  & 1, 99.8\%, 99.6\%, 99.5\% \\
        &$(0.08, 0.27, 0.30, 0.35)$& 6942  & 1, 99.6\%, 99.6\%, 99.5\%\\
        
        4.5& $(1/4, 1/4, 1/4, 1/4)$& 5557  & all 99.7\%\\
        &$(0.1, 0.2, 0.3, 0.4)$& 5926  &  1, 99.8\%, 99.6\%, 99.5\%  \\
        &$(0.08, 0.27, 0.30, 0.35)$&  5899 & 1, 99.6\%, 99.6\%, 99.5\%\\
    \hline
    \end{tabular}
    \label{table:LEADER-results-design-NI-LEADERassump}
\end{table}

\newpage
\appendix
\setcounter{equation}{0}
\renewcommand{\theequation}{A.\arabic{equation}}

\section{Appendix}

\subsection{Proof of Proposition \ref{prop:criterion-approxv2}}\label{app:criterion-approx-pfv2}

Let $A=(\begin{smallmatrix} 1 & -\pi \\ 0 & 1\end{smallmatrix})$. The joint distribution of $(\widetilde{D}_r-\pi \widetilde{D}, \widetilde{D})^\top$ is $N(A\vmu, A\Sigma A^\top)$ with
\[
A\vmu=\delta(1-\pi,1)^\top,\quad A\Sigma A^\top=w^{-1}\begin{pmatrix} \rho_r^{-1}-1+(1-\pi)^2 & 1-\pi \\ 1-\pi & 1 \end{pmatrix}.
\]

The conditional distribution of $\widetilde{D}_r-\pi\widetilde{D}$ given $\widetilde{D}$ is approximately $N(\widetilde{\mu}_r, \widetilde{\sigma}^{2(r)})$ with
\[
\widetilde{\mu}_r=(1-\pi)\widetilde{D}, \quad \widetilde{\sigma}^{2(r)}=\frac{1}{w}(\rho^{-1}_r-1).
\]
Then, 
\begin{equation}\label{eq:expression-condp}
\prob(\widetilde{D}_r\ge \pi\widetilde{D} \vert \widetilde{D}) = 
\prob(\widetilde{D}_r- \pi\widetilde{D}\ge 0 \vert \widetilde{D})  
\approx\Phi\left(\frac{\widetilde{\mu}_r}{\widetilde{\sigma}^{(r)}}\right)  
=\Phi\left(\frac{(1-\pi)\widetilde{D}\sqrt{w}}{\sqrt{\rho^{-1}_r-1}}\right)
\end{equation}

Let $Z=(\widetilde{D}-\delta)w^{1/2}$ denote the standardized quantity, which asymptotically follows the standard normal distribution. By (\ref{eq:nmrct}), substituting $\widetilde{D}w^{1/2}=Z+z_{1-\alpha}+z_{1-\beta}$ into (\ref{eq:expression-condp}),
\begin{equation}\label{eq:expression-condpv2}
\prob(\widetilde{D}_r\ge \pi\widetilde{D} \vert \widetilde{D}) 
\approx \Phi\left(\frac{(1-\pi)(Z+z_{1-\alpha}+z_{1-\beta})}{\sqrt{\rho^{-1}_r-1}}\right).
\end{equation}
Express the LHS of (\ref{eq:criterion}) as
\begin{equation} \label{eq:criterion-reorg}
    \frac{\prob(\widetilde{D}_r\ge\pi\widetilde{D}, T>z_{1-\alpha})}{\prob(T>z_{1-\alpha})}.
\end{equation}
On replacing $\widehat{\sd}(\widetilde{D})$ by $\sd(\widetilde{D})$, $\{T>z_{1-\alpha}\}$ is approximately $\{\widetilde{D}\sqrt{w}>z_{1-\alpha}\}$ which is equivalent to $\{Z>-z_{1-\beta}\}$ under the alternative hypothesis (i.e., $\delta>0$). Then the numerator of (\ref{eq:criterion-reorg}) is approximately the integration of $\prob(\widetilde{D}_r\ge \pi \widetilde{D} \vert \widetilde{D})$ w.r.t. $Z$, i.e.,
\[
\int_{u>-z_{1-\beta}}\Phi\left(\frac{(1-\pi) (u+z_{1-\alpha}+z_{1-\beta})}{\sqrt{\rho^{-1}_r-1}}\right) d\Phi(u)
\]
The denominator of (\ref{eq:criterion-reorg}) is approximately $\prob(Z>-z_{1-\beta})=1-\beta$. By substituting these into (\ref{eq:criterion-reorg}) yields (\ref{eq:criterion-approxv2}).

\subsection{Proof of Corollary \ref{co:cplowerbound}}\label{app:cplowerbound}
By (\ref{eq:nmrct}) and (\ref{eq: expression-in-hr}), 
\begin{equation}\label{eq:deno-rho}
\rho_r^{-1}-1=\frac{h_r}{h_r+1}\left(\frac{\tau^2(z_{1-\alpha}+z_{1-\beta})^2} {\delta^2}-\frac{h_r}{h_r+1}\right). 
\end{equation} 
For given $\delta$ and $\tau$, when $\frac{h_r}{h_r+1}=\frac{\tau^2}{2\delta^2}(z_{1-\alpha}+z_{1-\beta})^2<1$, (\ref{eq:deno-rho}) attains the maximum and (\ref{eq:cplowerbound-v2}) follows. Otherwise, the maximum value of (\ref{eq:deno-rho}) takes place when $\frac{h_r}{h_r+1}$ is sufficiently close to one, which gives (\ref{eq:cplowerbound-v3}).

\subsection{Proof of Corollary \ref{co:criterion-approxv2-fequal}}\label{app:criterion-approxv2-fequal}
When $f_r=R^{-1}$ for all $r$, $h_r={\tau^2n_0}/{(\Omega R)}$. By 
 (\ref{eq:SShomo}) and (\ref{app:cplowerbound}), 
\[\sqrt{\rho^{-1}_r-1}=\frac{\sqrt{R-1}\{\tau^2 (z_{1-\alpha}+z_{1-\beta})^2\}}{\delta^2R}.
\]
Then, the result follows. 

\subsection{Variance of the estimate of RMST}\label{sec:techDetailSur}

Let $T_{k}$ and $C_{k}$ denote the event time and censoring time of a subject group $k$, respectively. Assume that they are independent. Denote their associated survival functions by $S_k$ and $G_k$, respectively. We further denote the left continuous versions of $S_k$ and $G_k$ by $S_k^{-}$ and $G_k^{-}$, respectively. Denote the Kaplan-Meier estimators of $S_k$ and $G_k$, and the Nelson-Alaen estimator of the cumulative hazard function $\Lambda_k$ associated with $S_k$ by
$\widehat{S}_k(t)$, $\widehat{G}_k(t)$, and $\widehat{\Lambda}_k(t)$, respectively.

Recall $\widehat{\mu}_k=\int_0^{\eta}\widehat{S}_k(t)d(t)$. By \citet{pepe1989weighted}, the asymptotic variance of $\sqrt{n_k}\widehat{\mu}_k$ is
\begin{equation*}\label{eq: var-RMST-k}
\sigma_k^2=\int_{0}^{\eta} \frac{\{\int_t^{\eta}S_k(u)du\}^2}{S_k(t)G_k^{-}(t)} d\Lambda_k(t).
\end{equation*}
Its estimator $\widehat{\sigma}_k^{2}$ is obtained by replacing $S_k$, $G_k$ and $\Lambda_k$ with
$\widehat{S}_k(t)$, $\widehat{G}_k(t)$, and $\widehat{\Lambda}_k(t)$, respectively.

\bibliographystyle{chicago}
\bibliography{TR_rssMRCT}

\renewcommand{\thesection}{S\arabic{section}}
\renewcommand{\thefigure}{S\arabic{figure}}
\renewcommand{\thetable}{S\arabic{table}}
\setcounter{figure}{0}
\setcounter{table}{0}
\setcounter{section}{0}
\newpage
\section*{ Supplementary Materials for ``Consistency assessment and regional sample size calculation for MRCT under random effects model''}
\section{Tables and Figures}\label{sec:app-para}

\begin{table}[H]
    \centering
    \caption{Overall sample size $(n_0)$ solved from (\ref{eq:nmrct}) under various cases of $R$, $\delta$, $\tau/\delta$ and regional fractions, where $\alpha=0.025$, $\beta=0.1$, $\ell=1$ and $\Omega_r=2$ for all $r$}
    \begin{tabular}{ccc cc}
    \hline
    $R$& $(f_1,\ldots,f_R)$& $\tau/\delta$& $\delta=0.25$ &$\delta=0.5$\\
    \hline

        3& (1/3,1/3,1/3)& 0.2 & 392 & 98 \\
                        && 0.3 & 492 & 123 \\
                        && 0.4 & 765 & 192 \\
                        && 0.5 & 2704 & 676 \\
        \\
        
        &(0.2,0.3,0.5)& 0.2 & 399 & 100 \\
                        && 0.3 & 514 & 129 \\
                        && 0.4 & 828 & 207 \\
                        && 0.5 & 3055 & 764 \\
        \\                          
        4&(1/4,1/4,1/4,1/4)& 0.2 & 376 & 94 \\
                            && 0.3 & 441 & 111 \\
                            && 0.4 & 581 & 146 \\
                            && 0.5 & 980 & 245 \\
                            
        \\
        &(0.1,0.2,0.3,0.4)& 0.2 & 384 & 96 \\
                            && 0.3 & 464 & 116 \\
                            && 0.4 & 639 & 160 \\
                            && 0.5 & 1150 & 288 \\
    \hline    
    \end{tabular}
    \label{tab:ss-sensitive-omega1}
\end{table}

\begin{table}[H]
    \centering
    \caption{Sample size per group obtained by (\ref{eq:nmrct}), sample size per group of region 1, and the corresponding CP (in \%) obtained by (\ref{eq:criterion-approxv2}) under various configurations of $(f_1,\ldots,f_R)$ (with increasing $f_1$ and equal fractions in the remaining regions) given $\delta=0.25$, $\tau=0.1$, $\Omega_r=2$ for all $r$, and given $\ell=1$, $\pi=0.5$, $\alpha=0.025$ and $\beta=0.1$ }
    \begin{tabular}{*{5}{c}}
    \hline
    $R$ &  $(f_1, f_2, f_3)$& $n_0$& $n_0^{(1)}$& CP \\
    \hline
    3& (0.1,0.45,0.45)& 946& 95& 98.8\\
     & (0.3,0.35,0.35)& 768&  231& 97.5 \\
     & (0.5,0.25,0.25)& 817& 409& 97.0 \\
     \\
4 & (0.1,0.3,0.3,0.3)& 620& 62& 99.3\\
  &(0.3,7/30,7/30,7/30)& 584& 176& 97.8\\
  &(0.5,1/6,1/6,1/6)& 656& 328& 97.1\\ 
    \hline
    \end{tabular}
    \label{tab:CP-f1desc}
\end{table}

\begin{figure}[H]
    \centering
    \includegraphics[width=1\linewidth]{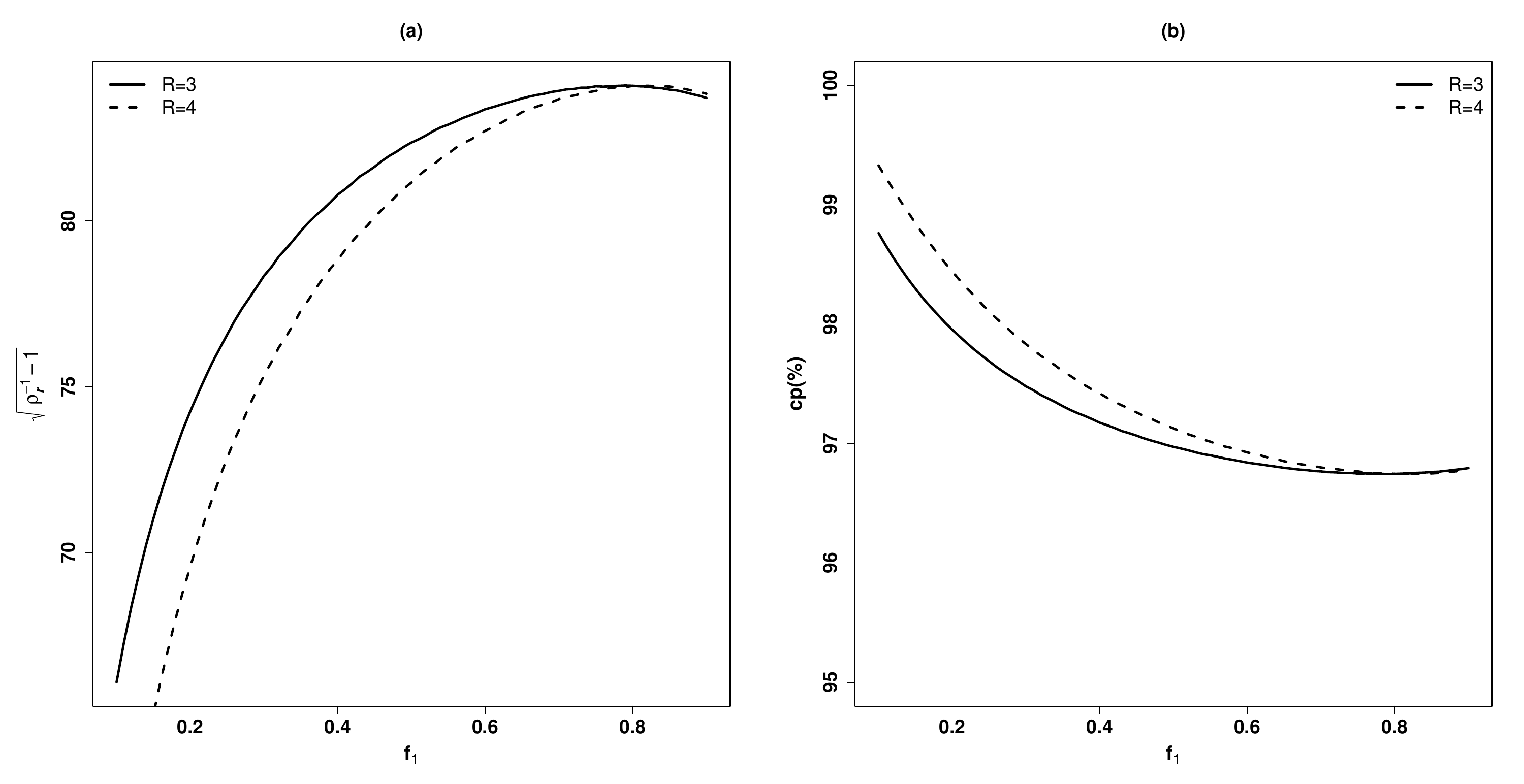}
    \caption{Values of $\sqrt{\rho_r^{-1}-1}$ and CP (\%) in (\ref{eq:criterion-approxv2}) as functions of $f_1$ when setting $f_2=\ldots=f_R=(1-f_1)/(R-1)$, $\alpha=0.025$, $\beta=0.1$, $\tau=0.1$, $\delta=0.25$, $\Omega_r=2$ for all $r$, $\ell=1$, and $\pi=0.5$}\label{fig:both-f1}
\end{figure}

\begin{table}[H]
    \centering
    \caption{Lower bounds of the CP (in \%) and possible solutions of $(n_0,f_1)$ under various combinations of $\alpha$, $\beta$ and $\tau/\delta$ given $\delta=0.25$, $\pi=0.5$, $\ell=1$ and $\Omega_1=2$, where the CP is obtained by (\ref{eq:cplowerbound-v2}) when $\tau/\delta=0.4$ and by (\ref{eq:cplowerbound-v3}) when $\tau/\delta=0.6$ as restricted by the condition in Corollary~\ref{co:cplowerbound} (n.a. stands for not available.)} 
    \begin{tabular}{*{5}{c}}
    \hline
    $\alpha$&  $\beta$& 
    $\tau/\delta$ & CP &  $(n_0,f_1)$\\
    \hline
    0.025& 0.1&   0.4& 96.7 & (10547, 0.1),\ (2110, 0.5)\\
         &     &  0.6& 84.1 & n.a.\\
         & 0.2&  0.4& 98.6 &(3376, 0.1),\ (676, 0.5)\\
         &     &  0.6& 86.9 & n.a.\\
     0.05& 0.1&  0.4& 97.6 &(4352, 0.1),\ (871, 0.5)\\
         &     &  0.6& 85.0 & n.a.\\
        &  0.2&  0.4& 99.2 &(1958, 0.1),\ (392, 0.5)\\
        &     &  0.6& 88.7 &  n.a.\\      
    \hline
    \end{tabular}
    \label{tab:CPlowerbound-prop2}
\end{table}

\begin{table}[H]
    \centering
    \caption{CPs (in \%) of (\ref{eq:criterion-approx-fequal}) for equal regional fraction and homogeneous variance under various combinations of $\alpha\in\{0.025, 0.05\}$, $\beta\in \{0.1, 0.2\}$, $R\in\{3, 4\}$ and $\tau/\delta\in\{0.4, 0.6\}$ with $\pi=0.5$ (n.a. stands for not available, which is due to the violation of condition $\tau/\delta<\sqrt{R}/(z_{1-\alpha}+z_{1-\beta})$.) }
    \begin{tabular}{*{5}{c}}
    \hline
    $\alpha$& $\beta$& $R$& $\tau/\delta$& CP\\
    \hline
    0.025& 0.1& 3& 0.4& 97.4\\
         &    &  & 0.6& n.a.\\ 
         &    &  4& 0.4& 98.1\\
         &    &  & 0.6& 84.5\\ 
    0.025& 0.2& 3& 0.4& 99.0\\
         &    &  & 0.6& 87.2\\ 
         &    & 4& 0.4& 99.3\\
         &    &  & 0.6& 89.1\\ 
     0.05& 0.1& 3& 0.4& 98.1\\
         &    &  & 0.6& n.a.\\ 
         &    & 4& 0.4& 98.6\\
         &    &  & 0.6& 86.8\\
     0.05& 0.2& 3& 0.4& 99.4\\
         &    &  & 0.6& 89.8\\ 
         &    & 4& 0.4& 99.6\\
         &    &  & 0.6& 91.5\\  
    \hline
    \end{tabular}
    \label{tab:CP-fequal}
\end{table}

\begin{table}[H]
\centering
\caption{Parameters of the distributions of survival endpoints representing early effect, late effect and crossing effect, where $\psi=10$ and 6 for early effect and late effect respectively. The resulting regional treatment effects and variances are provided in the last two columns. (Assume the censoring time follows the uniform distribution in $(0, 3\eta)$ with $\eta=80$.)}
\begin{tabular}{*{6}{c}}
\hline
{scenario} & {region} &  {($\lambda_0^{(r)},\gamma_0^{(r)}$)} &{($\lambda_1^{(r)},\gamma_1^{(r)}$)} &{$D_r$} &{($\sigma_1^{2(r)},\sigma_0^{2(r)}$)}\\
\hline
1 (early effect)&  1& (0.07,0.03)& (0.02,0.03)& 11.3 & (664,607) \\
            & 2& (0.07,0.04)& (0.03,0.04)& 7.2& (505,448) \\
            & 3& (0.07,0.05)& (0.04,0.05)& 4.4& (377,338) \\
            & 4& (0.07,0.06)& (0.05,0.06)& 2.5& (283,261) \\ 
            \\
                 
2 (late effect)& 1& (0.04,0.1)& (0.04,0.04)& 10.8& (498,113) \\
                & 2& (0.05,0.1)& (0.05,0.05)& 7.0& (367,113) \\
                & 3& (0.06,0.1)& (0.06,0.06)& 4.5& (273,112) \\
                & 4& (0.07,0.1)& (0.07,0.07)& 2.8& (208,111) \\ 
\\

 & &  {($\nu_0^{(r)}, \theta_0^{(r)}$)} &{($\nu_1^{(r)}, \theta_1^{(r)}$)} &  \\

3 (crossing)& 1& (1.6,30)& (1,50)& 13.1& (855,312) \\
            & 2& (1.6,40)& (1,60)& 9.0& (896,491) \\
            & 3& (1.6,50)& (1,70)& 5.3& (908,612) \\
            & 4& (1.6,60)& (1,80)& 2.3& (903,669) \\
 \hline
\end{tabular}
\label{table:result-3-sce1}
\end{table}

\begin{table}[H]
    \centering
    \caption{Results of subgroup analysis by region of the LEADER trial in event rate and HR (95\% CI) based on the fixed effects model and random effects model, where variance of regional treatment estimate ($\widehat{\sigma}^{2(r)}$) is provided. }
    \begin{tabular}{*{6}{c}}
    \hline
         &  Liraglutide&  Placebo&  & & \\
         region & (events/patients)& (events/patients)& HR (95\% CI)& $\widehat{\sigma}^{2(r)}$&  $\widetilde{\rm HR}$ (95\% CI)\\
    
    \hline     
         EU& 207/1639& 252/1657& 0.82(0.68,0.98)& 0.0087&0.84(0.71,1.00)\\
         NA& 212/1401& 216/1446& 1.01(0.84,1.22)&  0.0093& 0.93(0.78,1.10)\\
         Asia& 24/360& 37/351& 0.62(0.37,1.04) & 0.0656& 0.83(0.71,0.97)\\
         ROW& 165/1268& 189/1218& 0.83(0.68,1.03)&  0.0113& 0.85(0.71,1.01)\\
         Overall& 608/4683& 694/4672& 0.87(0.78,0.97)& -- &0.86(0.75,0.99)\\
    \hline
    \end{tabular}
    \label{table:LEADER-results}
\end{table}

\vskip 1in

\newpage

\begin{figure}[H]
    \centering
    \includegraphics[width=0.9\linewidth]{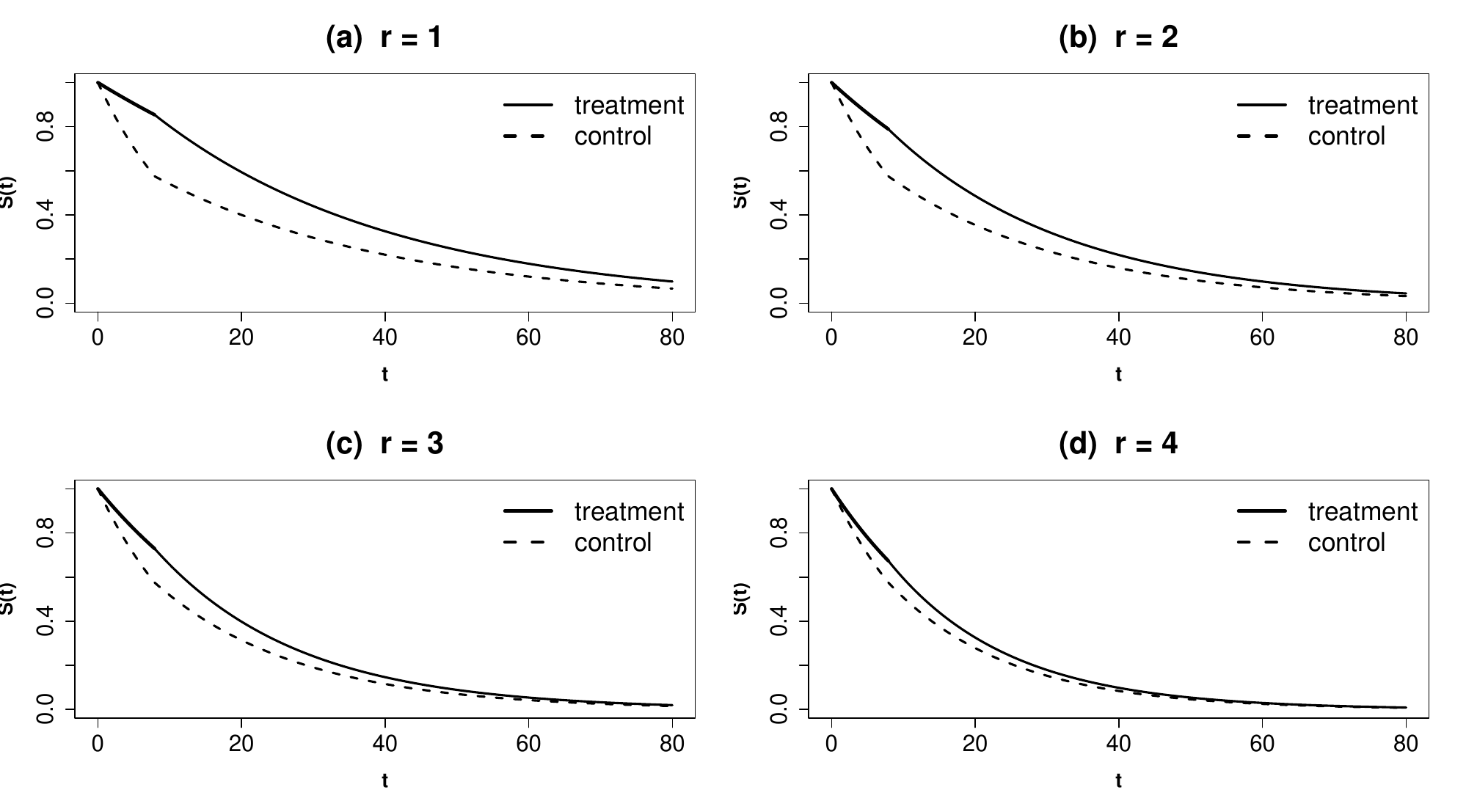}
    \caption{Survival functions that exhibits early effect in four regions, where the parameters are given in the upper panel of Table~\ref{table:result-3-sce1}}
    \label{fig:survnph-early-pw}
\end{figure}
\begin{figure}[H]
    \centering
    \includegraphics[width=0.9\linewidth]{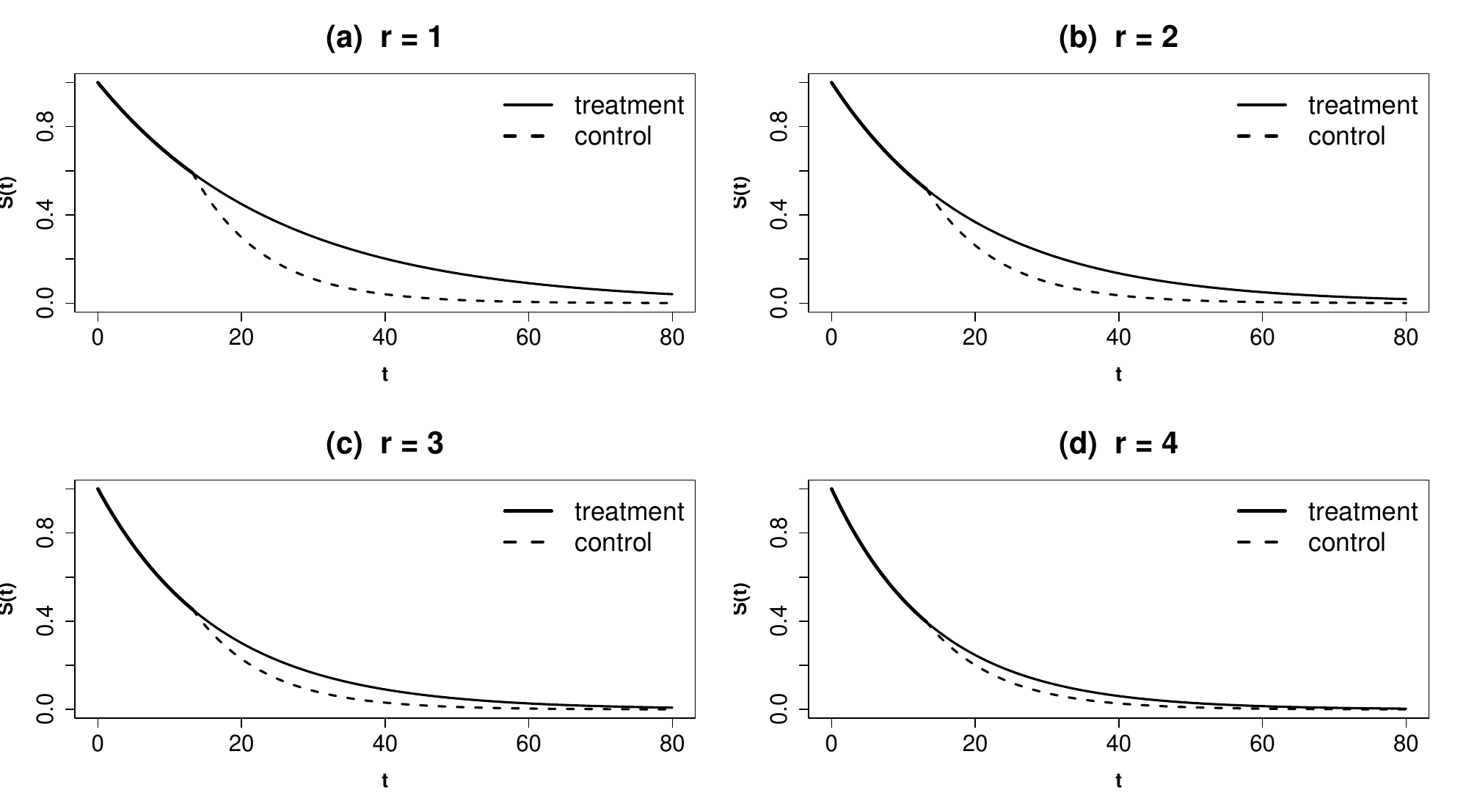}
    \caption{Survival functions that exhibits late effect in four regions, where the parameters are given in the middle panel of Table~\ref{table:result-3-sce1}}
    \label{fig:survnph-late-pw}
\end{figure}
\begin{figure}[H]
    \centering
    \includegraphics[width=0.9\linewidth]{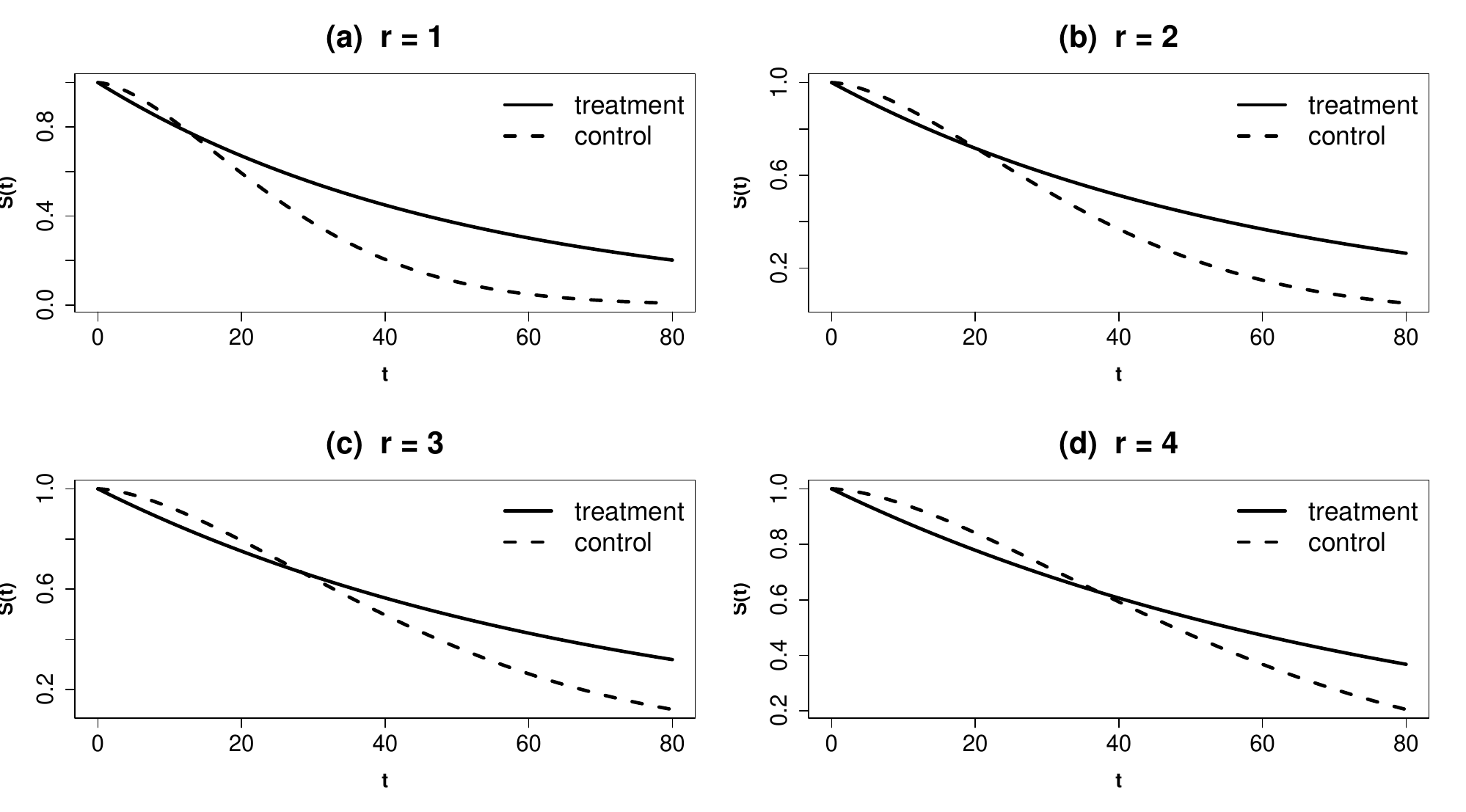}
    \caption{Survival functions that exhibits crossing effect in four regions, where the parameters are given in the lower panel of Table~\ref{table:result-3-sce1} }
    \label{fig:scenario9}
\end{figure}

\end{document}